# MEDEA: Automated Measure and On-Line Analysis in Astronomy and Astrophysics for Very Large Vision Machine

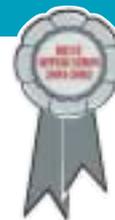

G. Iovane - Osservatorio di Capodimonte
Dipartimento di Ingegneria Informatica e Matematica Applicata –
Univ. of Salerno
I.N.F.N (Istituto Nazionale di Fisica Nucleare)

## THE CHALLENGE

The aim was to build a fully automated and real time system to perform image acquisition, reduction, and analysis, (for a 2k¥2k and 16k¥16k CCD camera) connected with an alert and "dispatcher" system to automatically inform researchers about the detection of relevant astrophysical event, like new Earth-like planets outside of the Solar system.

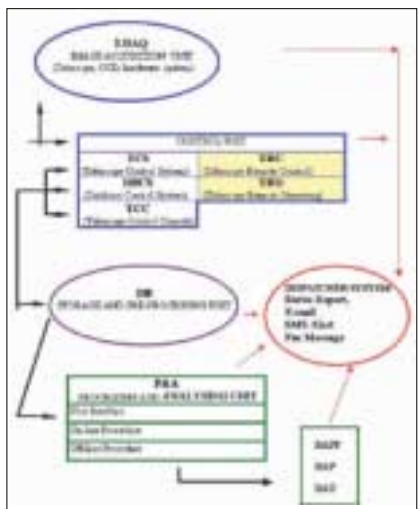

Fig.1: The skeleton of MEDEA environment.

## THE SOLUTION

MEDEA is a software architecture with a two trigger levels to automatically select relevant events and to detect and to recognize specially events. The information flow is controlled by a database system that is linked with a dispatcher system for informing and alerting people about special events.

## ABSTRACT

MEDEA is a software architecture to detect luminosity variations connected with the discovery of new planet outside the Solar System. Taking into account the enormous number of stars to monitor for our aim traditional approaches are very demanding in terms of computing time; here, the implementation of an automatic vision and decision system, which allows to perform an on-line discrimination of possible events by using two levels of trigger and a quasi-on-line data analysis, is presented. MEDEA becomes very profitable when dealing with large or deep surveys and provides to obtain soon results with its powerful data analysis tools.

## INTRODUCTION

During the last ten years, much attention has been devoted to the planet detection. The passage of the planet close to the line of sight of the observer implies a luminosity variation of the star that we are observing in the sky. MEDEA (Microlensing Experiment Data-Analysis Software for Events with Amplification) was implemented to give an answer to this point.

From a conceptual point of view the steps to process the data may be summarized as follows:

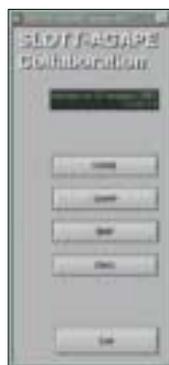

Fig.2: The main control panel that calls the secondary modules.

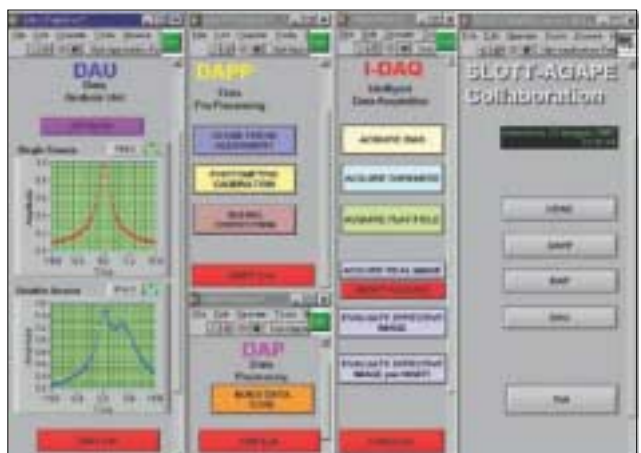

Fig.3: The full user interface; the sequence of the user operation is from right to left, infact on the right there is the main control panel, that calls the secondary ones: I-DAQ , DAPP, DAP, DAU.

1. Observation, measurement and image data acquisition;
2. Technical Image Processing (pre-reduction for bias, dark, pixel to pixel variation, cosmetics);
3. Evaluation of some quantities needed for the subsequent processing;
4. Astrometric (geometrical) alignment of images corresponding to different nights;
5. Photometric calibration of images corresponding to different nights;
6. PSF (Point Spread Function) correction of images corresponding to different nights;
7. First Trigger Level (selection of interesting luminosity variations through a peak detection algorithm);
8. Second Trigger Level (Selection of Planet Events);
9. Pilot Analysis (Quasi-On-Line);
10. Off - line Analysis.

## MEDEA ENVIRONMENT

MEDEA is structured as shown in Fig.1
i) The Image Data Acquisition (I-DAQ) Unit is responsible for the data acquisition and pre-reduction of data coming from images.
ii) The Control Unit, which thanks to Telescope Control System (TCS), controls the telescope by following the instructions provided by the DataBase Control System (DBCS) or by the user.
iii) The DataBase (DB) Unit is the intelligent part of the system; in fact, here we have the data storage and processing according to simulations or previous observations. Such unit is the natural link between observations and data analysis.
iv) The Processing and Analyzing (P&A) Unit is the platform where massive data analysis is performed. It consists of three main units: 1) Data Pre-Processing Unit (DAPP) for astrometric alignment, photometric calibra-tion, and PSF correction; Data Processing Unit (DAP) for peak detection of relevant luminosity variation and for removing unuseful objects; Data Analysis Unit (DAU) for the fits of light curve with different expected models, color correlation, $c^2$ test, Kolmogorov - Smirnov test.

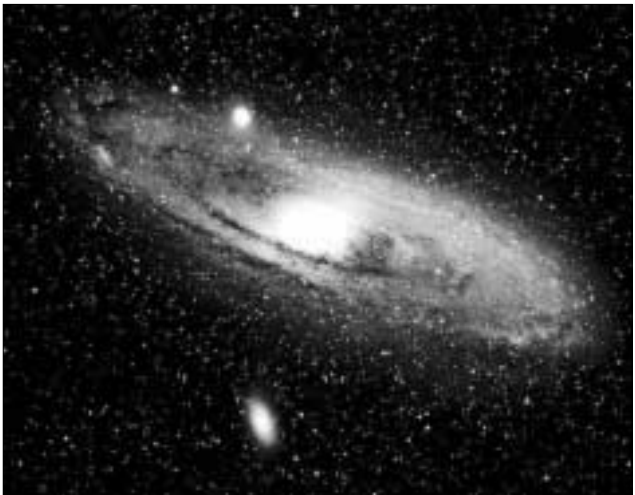

*Fig. 4: A typical field of view.*

The I-DAQ Unit performs steps 1-3. Steps 4-6 are taken core by the DAPP Unit. Both units are organized in a fully automatic (no human intervening) and on line fashion. Step 7 is the most relevant component of the DAP Unit and is composed by four sections: a) the peak detection procedure, b) the star detection and filtering algorithm, c) the cosmic rays filter, and d) the peak classification (single, double, multiple peak curve) routines. The DAP Unit runs in real time, this means that all data reduction is done just in the time between two following observations. The second trigger level corresponding to the step 8 is inside the DAU Unit. In particular, during this automated and real-time phase, we test also whether the measured luminosity curves are compatible with fixed models (like single point-like source and single point-like planet model, double point-like source and single point-like planet model, extended source with constant brightness and single point-like planet model) or not. The events that pass the first trigger level and are incompatible with the simple models included in the second trigger level are studied off line by using interactively procedures. This study is relevant to understand if those events are linked with complex events (like double planets, planetary system and so on) or variable stars or novae and supernovae. These tools are also included in the DAPP Unit, but they are off line and interactive.
v) The last Unit is the so-called Dispatcher, which automatically builds status report about the different phases of the data flow starting from the I-DAQ to the DAU unit. In particular, statistics, plots of data and events are produced and stored by this module. Moreover, in the oc-currence of special events (like an alert or failure of the system or a short event, for which a quick answer is needed) this unit can reach and alert people automatically via e-mail service and SMS (Short Message System).

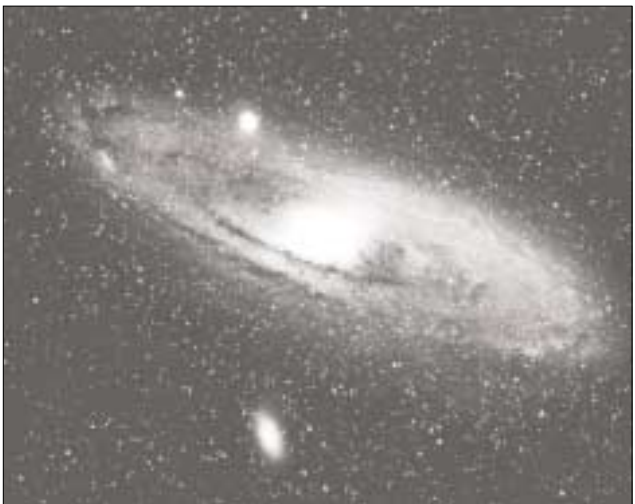

*Fig.5: The same field of Fig. 4, but in a different photometric condition.*

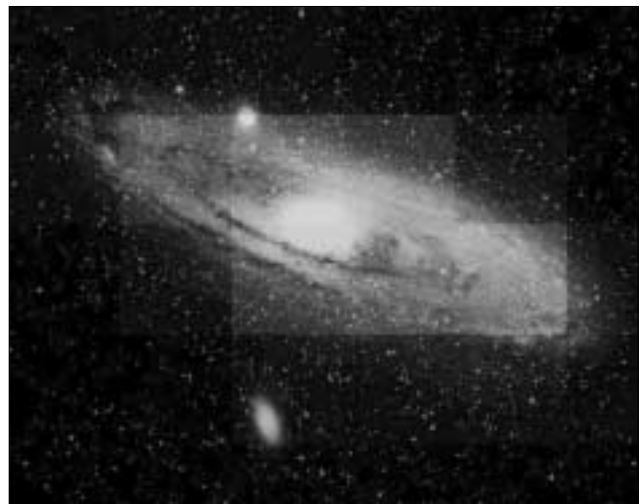

*Fig.6: The image after the local photometric correction (for our application the gradient on edges in not relevant).*

In Figs.2, 3 we find the user interface; in particular, in Fig.2 the main control panel is shown where the user selects the operations to perform (data acquisition, data pre-processing, data processing or data analysis), while Fig.3 shows all secondary panels.

### RESULTS OF DIFFERENT SOFTWARE MODULES

The first operation that we perform after the pre-reduction is the geometrical alignment of the images that is a translation plus a small rotation. Then we have the photometric alignment. Fig.4 and 5 show taken a typical field in different conditions, while in Fig.6 we find the local corrected image (the old Fig.5) for photometric effects. Before performing the comparison of the image, it is also necessary to operate a point spread function correction by using a morphing algorithm. Then thanks to the work in Fourier transformed space all resolved (large in terms of number of pixels) objects are removed (see Fig.7). At this step, by considering images at different time, we can build a light curve for each pixel. A typical light curve of a single source and single planet is shown in Fig.8, while in Fig.9 we find an event connected with a double source and single planet. To certify the events in the last operations we evaluate two statistical parameters: the c2 parameter and the Kolmogorov-Smirnov one; in addition we perform the correlation between the light curve of the same pixel, in different optical band to see whether the luminosity variation in different colors is the same.

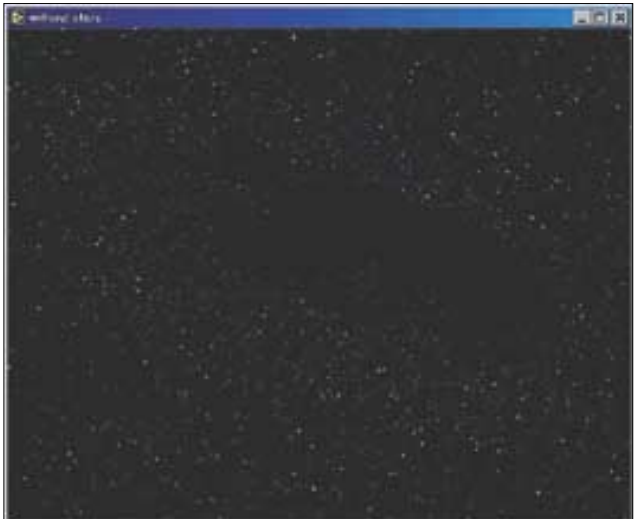

*Fig.7: The field after the filtering to remove extended objects (resolved stars, which are not interesting for our analysis).*

## CONCLUSION

MEDEA is designed for new automated planet searches. Most of the tools presented in this paper find possible applications in other fields, where pipelines and data mining procedures are called for a large amount of data. The procedures implemented in MEDEA environment and presented in Image Data Acquisition (I-DAQ), and Data pre-processing (DAPP) units were tested on simulated data and images, while the Data processing (DAP) and data analysis (DAU) units were tested on a small set of data collected with the 1.3 m McGraw-Hill telescope, at MDM observatory, Kitt Peak (USA) in the period from the end of September to the end of December 1999 by using the bulge of the Andromeda galaxy as target. The flexibility of the system is particularly useful to select the events to study on-line and off-line and by using automatic, semiautomatic, or assisted by researcher procedures. Thanks to this facility, more common events due to single point like source and planet, or extended source are analyzed on-line and with automatic procedures, while more complex events such as double sources, novae, supernovae, variable star or multiple planets system (planetary systems), double point like source and planet are studied off-line and respectively with automatic procedure (thanks to a database of simulated events) or with a traditional work of the researcher. In this way all the advantages of non-automatic procedure are kept, but in addition, common events are automatically studied so the researcher has only to certify them and thanks to the two trigger levels he spends his time only on physically relevant events. Moreover, thanks to the real time light curve monitoring and to the dispatcher implementation the detection of short events or events with a huge main peak and a secondary one near the first (like in planetary system) become possible. The follow-up observations of planet events, and the co-detection among different international collaborations become possible too. At the end the reduction of time to spend for data analysis after data taking is a good starting point to perform planet analysis on large field CCD cameras and deep surveys, so the powerful planet statistic really becomes a useful test to discriminate among different cosmological theory on the Universe and its evolution. Also thanks to MEDEA three new planet candidates are found from the international Community.

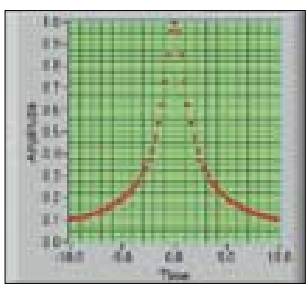

*Fig.8: A typical planet event corresponding to a single source.*

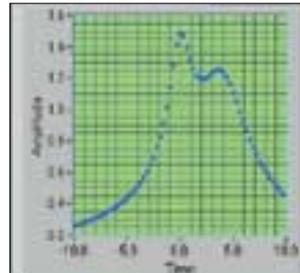

*Fig.9: A typical planet event corresponding to a double source.*

## ACKNOWLEDGEMENTS
The author wishes to thank SLOTT-AGAPE Collaboration for useful suggestions and comments about gravitational lensing, and A. Ambu, M. Quaglia, F. Selleri, A. Tamagnini from NI Italy for relevant software suggestions and comments.

**Prodotti utilizzati**
LabVIEW, Visione artificiale, SQL Toolkit, Internet Toolkit, Signal Processing Toolset